\RequirePackage{ifpdf}
\ifpdf % We are running pdfTeX in pdf mode
\documentclass[pdftex]{sigma}
\else
\documentclass{sigma}
\fi

\begin{document}

\numberwithin{equation}{section}
\newcommand{\R}{\mathbb{R}}
\newcommand{\C}{\mathbb{C}}
\newcommand{\tx}{\tilde{X}}
\newcommand{\tp}{\tilde{P}}
\newcommand{\talpha}{\tilde{\alpha}}
\newcommand{\tbeta}{\tilde{\beta}}
\newcommand{\tkappa}{\tilde{\kappa}}
\newcommand{\hx}{\hat{X}}
\newcommand{\hp}{\hat{P}}
\newcommand{\bx}{\bar{X}}
\newcommand{\bp}{\bar{P}}
\newcommand{\balpha}{\bar{\alpha}}
\newcommand{\bbeta}{\bar{\beta}}
\def\sech{\mathop{\rm sech}\nolimits}

\allowdisplaybreaks

\renewcommand{\PaperNumber}{016}

\FirstPageHeading

\ShortArticleName{Generalized Deformed Commutation Relations}

\ArticleName{Generalized Deformed Commutation Relations\\ with Nonzero Minimal
Uncertainties in Position\\ and/or Momentum and Applications\\ to Quantum Mechanics}

\Author{Christiane QUESNE~$^\dag$ and Volodymyr M. TKACHUK~$^\ddag$}

\AuthorNameForHeading{C. Quesne and V.M. Tkachuk}

\Address{$^\dag$~Physique Nucl\'eaire Th\'eorique et Physique Math\'ematique,  Universit\'e
Libre de Bruxelles, \\
$\phantom{^{\dag}}$~Campus de la Plaine CP229, Boulevard~du Triomphe, B-1050 Brussels,
Belgium}
\EmailD{\href{mailto:cquesne@ulb.ac.be}{cquesne@ulb.ac.be}} % E-mail address of First Author

\Address{$^\ddag$~Ivan Franko Lviv National University, Chair of Theoretical Physics,  \\
$\phantom{^{\ddag}}$~12 Drahomanov Str., Lviv UA-79005, Ukraine}
\EmailD{\href{mailto:tkachuk@ktf.franko.lviv.ua}{tkachuk@ktf.franko.lviv.ua}} % E-mail address of Second Author

\ArticleDates{Received November 22, 2006; Published online January 31, 2007}

\Abstract{Two generalizations of Kempf's quadratic canonical commutation relation in one
dimension are considered. The f\/irst one is the most general quadratic commutation
relation. The corresponding nonzero minimal uncertainties in position and momentum are
determined and the ef\/fect on the energy spectrum and eigenfunctions of the harmonic
oscillator in an electric f\/ield is studied. The second extension is a function-dependent
generalization of the simplest quadratic commutation relation with only a nonzero
minimal uncertainty in position. Such an uncertainty now becomes dependent on the
average position. With each function-dependent commutation relation we associate a
family of potentials whose spectrum can be exactly determined through supersymmetric
quantum mechanical and shape invariance techniques. Some representations of the
generalized Heisenberg algebras are proposed in terms of conventional position and
momentum operators $x$, $p$. The resulting Hamiltonians contain a contribution
proportional to $p^4$ and their $p$-dependent terms may also be functions of $x$. The
theory is illustrated by considering P\"oschl--Teller and Morse potentials.}


\Classification{37N20; 81R15}

\section{Introduction}

\looseness=1
During recent years, there has been much interest in studying theories characterized by a
mi\-ni\-mal observable length $\Delta X_0$. Some works \cite{gross, amati} in the context of
perturbative string theory have indeed suggested a generalized uncertainty principle, which,
for dimensionless operators, reads
\begin{gather}
  \Delta X \ge \frac{1}{2} \biggl( \frac{1}{\Delta P} + \beta \Delta P\biggr),  \label{eq:GUP}
\end{gather}
where $\beta$ is some very small positive parameter. Equation (\ref{eq:GUP}) leads to a
nonzero minimal uncertainty in position (or minimal length) $\Delta X_0 = \sqrt{\beta}$.
Another consequence of (\ref{eq:GUP}) is a mixing between UV and IR divergences. Both
properties have also emerged from many other studies in string theory and quantum gravity
(see, e.g., \cite{maggiore, connes, amelino, seiberg} and references quoted therein).

{\samepage
The generalized uncertainty principle (\ref{eq:GUP}) implies some modif\/ication of the
canonical commutation relations. Amongst several possibilities, one of the most promising and
simplest proposal~\cite{amati}
\begin{gather*}
  [X, P] = {\rm i} \big(1 + \beta P^2\big),
\end{gather*}
is based on the addition of some small quadratic correction. This line of approach has been
thoroughly investigated by Kempf \cite{kempf94a, hinrichsen, kempf95, kempf97}, who
considered more general quadratic corrections in $D$ dimensions and also suggested that the
absence of plane waves or momentum eigenvectors on generic curved spaces would give rise to
a minimal observable momentum $\Delta P_0$ \cite{kempf94b}.

}

Since the UV/IR mixing allows one to probe high-energy physics by low-energy one, it justif\/ies
the use of quantum mechanics in the presence of a minimal length. Investigating the inf\/luence
of the minimal length assumption on the energy spectrum of quantum systems has become
an interesting issue primarily for two reasons. First, this may help to set some upper bounds
on the value of the minimal length. In this connection, one may quote some studies of the
hydrogen atom \cite{brau99, benczik05, stetsko}, of the problem of a particle in a
gravitational quantum well \cite{brau06}, and of the application of the one-dimensional Dirac
oscillator to quark-gluon plasmas \cite{nouicer}. Furthermore, the classical limit has also
provided some interesting insights into some cosmological problems \cite{chang02a,
benczik02}. Second, it has been argued \cite{kempf97} that quantum mechanics with a
minimal length may also be useful to describe non-pointlike particles, such
as quasiparticles and various collective excitations in solids, or composite particles, e.g.,
hadrons. The formalism then provides us with an ef\/fective theory of
such systems in terms of a parameter $\Delta X_0$ related to the f\/inite size of the
particles.

Solving quantum mechanical problems with deformed canonical commutation relations,
however, usually turns out to be much more dif\/f\/icult than with conventional ones. This is
indeed the case when employing Kempf's quadratic commutation relations, so that only a few
examples have been considered in such a context.

In the simplest case where one considers a nonzero minimal uncertainty in position only,
several quantum mechanical systems have been successfully dealt with by solving the
corresponding Schr\"odinger equation in momentum space. Such problems include the one-
and $D$-dimensional harmonic oscillators~\cite{kempf95, chang02b}, for which exact
solutions have been provided, the hydrogen atom, for which perturbative \cite{brau99,
stetsko}, numerical \cite{benczik05}, or one-dimensional exact \cite{fityo06a} results have
been obtained, and the gravitational quantum well \cite{brau06}, which has been treated
perturbatively. The use of WKB approximation has also been tested \cite{fityo06b}.

In the more complicated case where one considers nonzero minimal uncertainties in both
position and momentum and one can therefore only resort to a generalized Fock space
(or corresponding Bargmann) representation~\cite{kempf94a, hinrichsen, kempf93}, we have
recently proposed an entirely dif\/ferent approach to the one-dimensional harmonic
oscillator problem~\cite{cq03}. It is based upon an extension to the deformed
commutation relation in hand of supersymmetric quantum mechanical (SUSYQM)
techniques~\cite{cooper}. When supplemented with shape invariance (SI) under
parameter translation~\cite{gendenshtein, dabrowska} or parameter
scaling~\cite{spiridonov92a, spiridonov92b, khare, barclay, lutzenko, loutsenko}, these are
known to provide a~very powerful method for exactly solving problems in standard quantum
mechanics. This is the case here too, since they have also allowed us~\cite{cq04a} to deal with
the one-dimensional harmonic oscillator in a uniform electric f\/ield and to algebraically rederive
the results of~\cite{kempf95, chang02b}.

The only relativistic problem that has been exactly solved so far in a deformed space with
minimal length is the Dirac oscillator, either in three dimensions using our extended SUSYQM
and SI method \cite{cq05}, or in one dimension by directly solving the dif\/ferential equation in
momentum representation \cite{nouicer}.

The purpose of the present paper is twofold: f\/irst, to propose some generalizations of
Kempf's quadratically-deformed canonical commutation relation in one dimension and to
analyze how such extensions af\/fect the uncertainties in position and momentum, and
second, to further illustrate the power of combined SUSYQM and SI techniques in solving
some eigenvalue problems corresponding to these new deformed commutation
relations.

At this stage, it should be stressed that our work is mainly aimed at applications to systems
of non-pointlike particles, especially in condensed-matter physics, where one-dimensional
systems may be experimentally produced. However, our ultimate goal would be to extend the
simple models presented here to more than one dimension, in which case the noncommutativity
of position coordinates (and possibly also that of momentum ones) would become of utmost
importance.

The paper is organized as follows. The most general quadratic canonical commutation
relation with nonzero minimal uncertainties in both position and momentum is considered
in Section~2. Then a function-dependent generalization of the quadratic commutation
relation with only a~nonzero minimal uncertainty in position is reviewed in Section~3.
Finally, in Section~4, we summarize our results and sketch some of their possible
applications to a variety of problems.

\section{Generalized quadratic commutation relation}

\subsection{Commutation relation and uncertainty relation}

The most general quadratically-deformed canonical commutation relation can be written as
\begin{gather}
  [X, P] = {\rm i}(1 + \alpha X^2 + \beta P^2 + \kappa XP + \kappa^* PX),
  \label{eq:quad-com}
\end{gather}
where $\alpha, \beta \in \R$ and $\kappa \in \C$ are assumed to be very small
parameters (i.e., $|\alpha|, |\beta|, |\kappa| \ll 1$), while~$X$ and~$P$ are dimensionless
operators. Throughout this paper, we use units wherein $\hbar=1$ (as well as $\ell = m =
1$, where $\ell$ and $m$ denote the characteristic length and the particle mass for the
quantum mechanical system under consideration).

{\sloppy
Equation (\ref{eq:quad-com}) admits two important special cases: choosing $\kappa=0$
and $\alpha, \beta > 0$ leads to Kempf's commutation relation characterized by nonzero
minimal uncertainties $\Delta X_0 = \sqrt{\beta/(1 - \alpha\beta)}$ and $\Delta P_0 =
\sqrt{\alpha/(1 - \alpha\beta)}$\,\footnote{It should be stressed that the positivity of
$\alpha$ and $\beta$ is essential to get nonvanishing values for $\Delta X_0$ and
$\Delta P_0$.}, whereas selecting $\alpha=\beta=0$ gives rise to the $q$-deformed
Heisenberg algebra $qXP - q^*PX = {\rm i}$ with $q \equiv 1 - {\rm i}\kappa$
\cite{pillin}.

}

In the general case, on setting $\kappa = \kappa_1 + {\rm i} \kappa_2$, $\kappa_1,
\kappa_2 \in \R$, equation (\ref{eq:quad-com}) may be rewritten as
\begin{gather}
  [\tx, \tp] = {\rm i}[1 + \talpha \tx^2 + \tbeta \tp^2 + \tkappa_1(\tx \tp + \tp \tx)]
  \label{eq:quad-com-tilde}
\end{gather}
in terms of some rescaled operators and parameters, $\tx = X \sqrt{1 + \kappa_2}$,
$\tp = P \sqrt{1 + \kappa_2}$, $\talpha = \alpha/(1 + \kappa_2)$, $\tbeta =
\beta/(1 + \kappa_2)$ and $\tkappa_1 = \kappa_1/(1 + \kappa_2)$. This simple
property allows us to restrict ourselves to real values of $\kappa$. Hence, in the
remainder of this section, we will inquire into the inf\/luence of an additional term ${\rm i}
\kappa (XP + PX)$, with $\kappa \in \R$, on the right-hand side of Kempf's deformed
commutation relation
$[X, P] = {\rm i}(1 + \alpha X^2 + \beta P^2)$ with $\alpha, \beta \in \R^+$.

On performing the rotation
\begin{gather*}
  X = X' \cos\varphi + P' \sin\varphi, \qquad P = - X' \sin\varphi + P' \cos\varphi,
\end{gather*}
such a generalized deformed commutation relation can be rewritten as
\begin{gather}
  [X', P'] = {\rm i}[1 + \alpha' X^{\prime2} + \beta' P^{\prime2} + \kappa' (X'P' + P'X')],
  \label{eq:quad-com-prime}
\end{gather}
where
\begin{gather}
  \alpha'  = \frac{1}{2}(\alpha + \beta) + \frac{1}{2}(\alpha - \beta) \cos2\varphi -
       \kappa \sin2\varphi, \nonumber\\
  \beta' = \frac{1}{2}(\alpha + \beta) - \frac{1}{2}(\alpha - \beta) \cos2\varphi +
       \kappa \sin2\varphi, \nonumber \\
  \kappa' = \frac{1}{2}(\alpha - \beta) \sin2\varphi  + \kappa \cos2\varphi.
       \label{eq:alpha-prime}
\end{gather}

Equation (\ref{eq:quad-com-prime}) can be simplif\/ied by choosing $\kappa'=0$. Here we
have to distinguish between $\alpha \ne \beta$ and $\alpha = \beta$. In the former case
\begin{gather*}
  \tan2\varphi = - \frac{2\kappa}{\alpha - \beta},
\end{gather*}
where we may restrict $\varphi$ to the interval $- {\pi}/{4} < \varphi <
{\pi}/{4}$. As a consequence
\begin{gather}
  \cos2\varphi = \frac{|\alpha - \beta|}{\delta}, \qquad \sin2\varphi = -
  \frac{2\sigma \kappa}{\delta}, \qquad \delta \equiv \sqrt{(\alpha - \beta)^2 + 4
  \kappa^2}, \qquad \sigma \equiv \frac{(\alpha - \beta)}{|\alpha - \beta|}.
  \label{eq:varphi}
\end{gather}
In the latter case, $\cos2\varphi = 0$ may be achieved by choosing $\varphi =
{\pi}/{4}$, so that
\begin{gather}
  \alpha' = \alpha - \kappa, \qquad \beta' = \beta + \kappa. \label{eq:alpha-prime-bis}
\end{gather}

With such choices for $\varphi$, the transformed operators $X'$, $P'$ satisfy a
commutation relation similar to that of Kempf provided $\alpha'$ and $\beta'$ are
positive. From (\ref{eq:alpha-prime}), this amounts to the two conditions
\begin{gather*}
  \alpha + \beta > 0, \qquad \biggl|\frac{1}{2}(\alpha - \beta) \cos2\varphi - \kappa
  \sin2\varphi \biggr| < \frac{1}{2} (\alpha + \beta).
\end{gather*}
If $\alpha \ne \beta$, equation (\ref{eq:varphi}) allows us to transform the second
condition into $\kappa^2 < \alpha \beta$, which can only be valid if $\alpha \beta > 0$.
On combining the results, we are led to the restrictions
\begin{gather}
  \alpha > 0, \qquad \beta > 0, \qquad |\kappa| < \sqrt{\alpha \beta}.
  \label{eq:cond-alpha}
\end{gather}
It is straightforward to see that equation (\ref{eq:cond-alpha}) remains true for $\alpha
= \beta$. We therefore conclude that the generalized commutation relation
(\ref{eq:quad-com}) with $\kappa \in \R$ is equivalent to Kempf's one up to a rotation
of angle $- {\pi}/{4} < \varphi \le {\pi}/{4}$ if and only if equation
(\ref{eq:cond-alpha}) is fulf\/illed.

We now plan to f\/ind the minimal uncertainties $\Delta X_0$ and $\Delta P_0$ corresponding
to (\ref{eq:quad-com}), (\ref{eq:cond-alpha}), and generalizing those obtained by Kempf.

The uncertainty relation associated with (\ref{eq:quad-com}) for real values of $\kappa$
can be written as
\begin{gather}
  \Delta X \Delta P \ge \frac{1}{2} \bigl|1 + \gamma + \alpha (\Delta X)^2 + \beta
  (\Delta P)^2 + \kappa \langle \hx \hp + \hp \hx\rangle\bigr|, \label{eq:uncert1}
\end{gather}
where $\hx \equiv X - \langle X\rangle$, $\hp \equiv P - \langle P\rangle$, and
$\gamma \equiv \alpha \langle X\rangle^2 + \beta \langle P\rangle^2 + 2\kappa
\langle X\rangle \langle P\rangle$. From the inequality $|\langle AB + BA\rangle| \le 2
\sqrt{\langle A^2\rangle \langle B^2\rangle}$ valid for any two Hermitian operators $A$,
$B$, it follows that $|\langle \hx \hp + \hp \hx\rangle| \le 2 \Delta X \Delta P$.
Furthermore, conditions (\ref{eq:cond-alpha}) imply that $\gamma \ge 0$, hence $1 +
\gamma +
\alpha (\Delta X)^2 + \beta (\Delta P)^2 > 0$. Equation (\ref{eq:uncert1}) may
therefore be transformed into
\begin{gather}
  \Delta X \Delta P \ge \frac{1}{2}[1 + \gamma + \alpha (\Delta X)^2 + \beta
  (\Delta P)^2 - 2|\kappa| \Delta X \Delta P], \label{eq:uncert1bis}
\end{gather}
where we have used property (\ref{eq:cond-alpha}) again to drop the absolute value on
the right-hand side. It is now straightforward to rewrite (\ref{eq:uncert1bis}) as
\begin{gather}
  \Delta \bx \Delta \bp \ge \frac{1}{2}[1 + \balpha (\Delta \bx)^2 + \bbeta
  (\Delta \bp)^2] \label{eq:uncert1ter}
\end{gather}
with
\begin{gather*}
  \Delta \bx \equiv \sqrt{\frac{1 + |\kappa|}{1 + \gamma}}\, \Delta X, \qquad
  \Delta \bp \equiv \sqrt{\frac{1 + |\kappa|}{1 + \gamma}}\, \Delta P, \qquad
  \balpha \equiv \frac{\alpha}{1 + |\kappa|}, \qquad \bbeta \equiv \frac{\beta}{1 +
  |\kappa|}.
\end{gather*}

Since the last inequality (\ref{eq:uncert1ter}) is of the same type as that considered by
Kempf, we know that there exist nonzero minimal values $\Delta \bx_0 = \sqrt{\bbeta/
(1 - \balpha \bbeta)}$, $\Delta \bp_0 = \sqrt{\balpha/(1 - \balpha \bbeta)}$ of
$\Delta \bx$ and~$\Delta \bp$. From this, we infer that there also exist nonzero minimal
values of $\Delta X$ and $\Delta P$, given by
\begin{gather*}
  \Delta X_{\rm min} = \sqrt{\frac{\beta (1 + \gamma)}{(1 + |\kappa|)^2 - \alpha
  \beta}}, \qquad \Delta P_{\rm min} = \sqrt{\frac{\alpha (1 + \gamma)}{(1 +
  |\kappa|)^2 - \alpha \beta}},
\end{gather*}
so that the absolutely smallest uncertainties in $X$ and $P$ are
\begin{gather}
  \Delta X_0 = \sqrt{\frac{\beta}{(1 + |\kappa|)^2 - \alpha \beta}}, \qquad \Delta P_0 =
  \sqrt{\frac{\alpha}{(1 + |\kappa|)^2 - \alpha \beta}}. \label{eq:minuncert1}
\end{gather}

Had we considered equation (\ref{eq:quad-com}) for complex values of $\kappa$ (and
positive values of $\alpha$, $\beta$), equation (\ref{eq:minuncert1}) would have been
valid for $\Delta \tx_0$, $\Delta \tp_0$, $\talpha$, $\tbeta$, and $\tkappa_1$,
provided $|\tkappa_1| < \sqrt{\talpha \tbeta}$. Rescaling the operators and parameters,
as explained below equation (\ref{eq:quad-com-tilde}), would then have led us to minimal
uncertainties given by
\begin{gather}
  \Delta X_0 = \sqrt{\frac{\beta}{(1 + \kappa_2 + |\kappa_1|)^2 - \alpha \beta}}, \qquad
  \Delta P_0 = \sqrt{\frac{\alpha}{(1 + \kappa_2 + |\kappa_1|)^2 - \alpha \beta}},
  \label{eq:minuncert1bis}
\end{gather}
provided $|\kappa_1| < \sqrt{\alpha \beta}$. Observe that setting $\kappa_1 =
\kappa_2 = 0$ in (\ref{eq:minuncert1bis}) gives back Kempf's results, as it should be.

\subsection[Application to the harmonic oscillator in an electric field]{Application
to the harmonic oscillator in an electric f\/ield}

In the present subsection, we plan to study the inf\/luence of the $\kappa$-dependent
terms in equation~(\ref{eq:quad-com}) on the one-dimensional quantum mechanical
systems considered in \cite{cq03, cq04a} under the assumptions that $\kappa \in \R$
and conditions (\ref{eq:cond-alpha}) are satisf\/ied.

Since the harmonic oscillator Hamiltonian satisf\/ies the relation $H = ({1}/{2})(P^2 +
X^2) = ({1}/{2})(P^{\prime2} + X^{\prime2})$, where $X'$ and $P'$ fulf\/il equation
(\ref{eq:quad-com-prime}) with $\kappa' = 0$, its spectrum is independent of the
presence of $\kappa$-dependent terms in (\ref{eq:quad-com}).

This is not the case, however, for the harmonic oscillator in a uniform electric f\/ield $\cal
E$, since its Hamiltonian becomes
\begin{gather}
  H = \frac{1}{2}(P^2 + X^2) - {\cal E} X = \frac{1}{2}(P^{\prime2} + X^{\prime2}) -
  {\cal E}(X' \cos\varphi + P' \sin\varphi), \label{eq:H-field}
\end{gather}
where there is an additional term proportional to $P'$. To take care of this change, let us
facto\-rize~$H$ as
\begin{gather}
  H = B^+(g,s,r,\nu) B^-(g,s,r,\nu) + \epsilon_0, \label{eq:fact-H}
\end{gather}
where
\begin{gather}
  B^{\pm}(g,s,r,\nu) = \frac{1}{\sqrt{2}}(\mp {\rm i}gP' + sX' +r \mp {\rm i}\nu)
  \label{eq:B-field}
\end{gather}
with $s$, $g$, $r$, $\nu$, $\epsilon_0 \in \R$ and $s$, $g >0$. Observe that $\nu$ is a
new parameter not appearing in \cite{cq04a}. From (\ref{eq:H-field})--(\ref{eq:B-field}),
we get
\begin{gather*}
  g  = sk,\qquad  s  = \frac{1}{\sqrt{1 - \alpha' k}},\qquad
  k  = \frac{1}{2}\left(\beta' - \alpha' + \sqrt{1 + \frac{1}{4}(\beta' - \alpha')^2}\right),
  \\
  r = - \frac{{\cal E} \cos\varphi}{s},\qquad  \nu = - \frac{{\cal E} \sin\varphi}{g},\qquad
 \epsilon_0  = \frac{1}{2}\big(gs - r^2 - \nu^2\big),
\end{gather*}
in terms of the transformed parameters $\alpha'$, $\beta'$, given either in
(\ref{eq:alpha-prime}) and (\ref{eq:varphi}) or in (\ref{eq:alpha-prime-bis}).

More generally, $H$ is the f\/irst member $H_0$ of a hierarchy of Hamiltonians
\begin{gather*}
  H_i = B^+(g_i, s_i, r_i, \nu_i) B^-(g_i, s_i, r_i, \nu_i) + \sum_{j=0}^i \epsilon_j,
  \qquad i=0, 1, 2, \ldots,
\end{gather*}
satisfying the SI condition
\begin{gather}
B^-(g_i, s_i, r_i, \nu_i) B^+(g_i, s_i, r_i, \nu_i) \nonumber
\\\qquad
 = B^+(g_{i+1}, s_{i+1}, r_{i+1},
  \nu_{i+1}) B^-(g_{i+1}, s_{i+1}, r_{i+1}, \nu_{i+1}) + \epsilon_{i+1},
\label{eq:SI}
\end{gather}
where $i=0, 1, 2,\ldots$, $g_0=g$, $s_0=s$, $r_0=r$, and $\nu_0=\nu$. The solution
to such a condition is similar to that carried out in \cite{cq04a}, so that we only state
here the results
\begin{gather*}
  g_i  = g q^{i/2} \frac{1 + tq^{-i}}{1 + t}, \qquad
  s_i = s q^{i/2} \frac{1 - tq^{-i}}{1 - t}, \qquad
  r_i = r q^{-i/2} \frac{1 - t}{1 - tq^{-i}},
  \\
  \nu_i = \nu q^{-i/2} \frac{1 + t}{1 + tq^{-i}}, \qquad
  \epsilon_{i+1} =\frac{1}{2}\big(g_i s_i + g_{i+1} s_{i+1}
  + r_i^2 - r_{i+1}^2 + \nu_i^2 - \nu_{i+1}^2\big),
\end{gather*}
where
\begin{gather*}
  q \equiv \frac{1 + \sqrt{\alpha'\beta'}}{1 - \sqrt{\alpha'\beta'}}, \qquad t
  \equiv  \frac{g - \gamma s}{g + \gamma s}, \qquad
  \gamma \equiv  \sqrt{\frac{\beta'}{\alpha'}}.
\end{gather*}

The energy spectrum of $H$ now reads
\begin{gather*}
  E_n({\cal E}) = \sum_{i=0}^n \epsilon_i = E_n(0) + \Delta E^{(1)}_n({\cal E})
  + \Delta E^{(2)}_n({\cal E}),
\end{gather*}
where $E_n(0)$ are the harmonic oscillator Hamiltonian eigenvalues in the absence of
electric f\/ield (see equation (2.25) of \cite{cq04a}), while
\begin{gather*}
  \Delta E^{(1)}_n({\cal E})  = - \frac{2\gamma^2 {\cal E}^2 \cos^2\varphi}{u^2}
       q^{-n} (1 - tq^{-n})^{-2} = - \frac{1}{2} K^2 q^n z_n^2, \\
  \Delta E^{(2)}_n({\cal E})  = - \frac{2{\cal E}^2 \sin^2\varphi}{u^2}
       q^{-n} (1 + tq^{-n})^{-2} = - \frac{1}{2} K^2 q^n w_n^2
\end{gather*}
are two correction terms due to the electric f\/ield. Here $u \equiv g + \gamma s$, $K
\equiv u \sqrt{(q+1)/(4\gamma)}$, $z_i \equiv - r_i/(K q^{i/2})$, and $w_i \equiv -
\nu_i/(K q^{i/2})$. The f\/irst correction term was already present in \cite{cq04a} (note,
however, the replacement of $\cal E$ by ${\cal E} \cos\varphi$ and of $\alpha$,
$\beta$ by $\alpha'$, $\beta'$), whereas the second one is new. Both of them are
$n$-dependent, negative and increasing (from $- ({1}/{2}) K^2 z^2$ or $- ({1}/{2})
K^2 w^2$ to 0) when $n$ goes from 0 to $\infty$.

As in \cite{cq04a}, the corresponding eigenvectors can be written in the Bargmann
representation associated with some $q$-boson creation and annihilation operators
$b^+$, $b$, such that $b b^+ - q b^+ b = 1$. This is based upon the observation that
$B^-(g_i, s_i, r_i, \nu_i)$ can be expressed as $B^-(g_i, s_i, r_i, \nu_i) = (1/\sqrt{2})
K q^{i/2} (b - t_i b^+ - z_i - {\rm i} w_i)$ and that a similar result applies to the adjoint
operator $B^+(g_i, s_i, r_i, \nu_i)$. The ground-state wavefunction $\psi_0(q, t, z, w;
\xi)$, where $\xi \in \C$ represents $b^+$, is obtained in the same form as before (see
equation (2.59) of \cite{cq04a}), except for the substitution of the complex parameter
$z + {\rm i}w$ for the real one $z$. The excited-state wavefunctions $\psi_n(q, t, z, w;
\xi) \propto P_n(q, t, z, w; \xi) \psi_0(q, t_n, z_n, w_n; \xi)$ (with $t_n = q^n t$)
undergo a more profound transformation since, on the right-hand side of the recursion
relation for the $n$th-degree polynomials $P_n(q, t, z, w; \xi)$ given in equation (2.69)
of \cite{cq04a}, $z$ is replaced by $z - {\rm i}w$ whereas $z_{n+1}$ is changed into
$z_{n+1} + {\rm i} w_{n+1}$. For $n=1$ and 2, for instance, we obtain
\begin{gather*}
  P_1(q, t, z, w; \xi) = (1 - t^2 q^{-1})[\xi - (1 - t q^{-1})^{-1} z + {\rm i} (1 + t
        q^{-1})^{-1} w],
        \\
P_2(q, t, z, w; \xi)  = (1 - t^2 q^{-3})\{(1 - t^2 q^{-1}) \xi^2 - [2]_q (1 - t^2
q^{-1})[(1 - t q^{-2})^{-1} q^{-1} z
        \\ \qquad
{}- {\rm i} (1 + t q^{-2})^{-1} q^{-1} w] - t + (1 - t)(1 + t q^{-1})(1 - t
        q^{-2})^{-2} q^{-1} z^2
        \\ \qquad
 {}- (1 + t)(1 \!-\! t q^{-1})(1 + t q^{-2})^{-2} q^{-1} w^2 \!-\! 2{\rm i} (1 \!-\! t^2
        q^{-1}) [(1 \!-\! t q^{-2})(1 + t q^{-2})]^{-1}q^{-1} z w\},
\end{gather*}
respectively.

It is worth stressing that in contrast with the energy eigenvalues, which could be derived
from those obtained in \cite{loutsenko}, the corresponding eigenfunctions are an entirely
new and nontrivial result (see section 2.4 of \cite{cq04a} for a detailed comparison
between our approach and that of \cite{loutsenko}).

\section{Function-dependent commutation relation}

\subsection{Commutation relation and uncertainty relation}

Let us consider the commutation relation
\begin{gather}
  [f(X), P] = {\rm i}[f'(X) + \beta P^2],  \label{eq:f-com}
\end{gather}
where $f(X)$ is some dif\/ferentiable, real-valued function of a generalized position operator
$X$, $f'(X)$ denotes its derivative with respect to $X$, and $\beta \in \R^+$ is some
very small parameter (i.e., $\beta \ll 1$). For $\beta = 0$, equation (\ref{eq:f-com})
reduces to the relation $[f(x), p] = {\rm i} f'(x)$, satisf\/ied by the conventional position and
momentum operators $x$ and $p = - {\rm i} d/dx$, characterized by $[x, p] = {\rm i}$.
Furthermore, for $\beta \ne 0$, it may be considered as a generalization of Kempf's
quadratic commutation relation (\ref{eq:quad-com}) with $\alpha = \kappa = 0$
(leading to a nonzero minimal uncertainty $\Delta X_0 = \sqrt{\beta}$ in position only)
since the latter may be retrieved by choosing $f(X) = X$.

For some reasons that will be explained in Section 3.3, we are going to restrict
ourselves here to functions $f(X)$ such that
\begin{gather}
  f'(X) = a f^2(X) + b f(X) + c  \label{eq:f'}
\end{gather}
for some choice of real constants $a$, $b$, and $c$. This includes the Kempf's case for
which $a=b=0$ and $c=1$. Some other interesting cases to be considered later on are
\begin{gather}
  f(X)  = - e^{-X},\qquad f'(X) = e^{-X},\qquad a = 0,\qquad b = -1,\qquad c = 0, \label{eq:f-ex1} \\
  f(X)  = \tanh X,\qquad f'(X) = \sech^2 X,\qquad a = -1,\qquad b = 0,\qquad c = 1, \label{eq:f-ex2} \\
  f(X)  = \tan X,\qquad f'(X) = \sec^2 X,\qquad a = 1,\qquad b = 0,\qquad c =1.
      \label{eq:f-ex3}
\end{gather}
Observe that in all these examples, $X$ is the lowest-order term in an expansion of
$f(X)$ into powers of $X$.

To interpret the generalized commutation relation (\ref{eq:f-com}) from a physical
viewpoint, it is necessary to derive an approximate expression for $[X, P]$. For such a
purpose, let us restrict ourselves to those states for which $(\Delta X)^2 = \langle
\hx^2\rangle$ (where $\hx = X - \langle X\rangle$) is very small and of the order
of~$\beta$. On expanding $f(X)$ and $f'(X)$ around $\langle X\rangle$ and inserting such
expressions in (\ref{eq:f-com}), we arrive at the relation
  \begin{gather}
  [\hx, P]  = {\rm i} \biggl(1 + \frac{f''(\langle X\rangle)}{f'(\langle X\rangle)} \hx +
      \frac{1}{2} \frac{f'''(\langle X\rangle)}{f'(\langle X\rangle)} \hx^2 + \cdots +
      \frac{\beta}{f'(\langle X\rangle)} P^2\biggr) - \frac{1}{2} \frac{f''(\langle
      X\rangle)}{f'(\langle X\rangle)}\label{eq:f-com-bis}
  \\ \phantom{[\hx, P]  =}{}
 {}\times (\hx [\hx, P] \!+\! [\hx, P] \hx) - \frac{1}{6} \frac{f'''(\langle
      X\rangle)}{f'(\langle X\rangle)} (\hx^2 [\hx, P] \!+\! \hx [\hx, P] \hx
\!+\! [\hx, P] \hx^2) \!+\! \cdots. \nonumber
  \end{gather}
Expanding next $[\hx, P]$ into powers of $\hx$,
\begin{gather*}
  [\hx, P] = [\hx, P]_{(0)} + [\hx, P]_{(1)} \hx + [\hx, P]_{(2)} \hx^2 + \cdots,
\end{gather*}
on both sides of (\ref{eq:f-com-bis}) and equating successively the $k$th-order terms in
$\hx$ up to $k=2$ leads to the results
\begin{gather*}
  [\hx, P]_{(0)} = {\rm i}, \qquad [\hx, P]_{(1)} \hx = 0, \qquad [\hx, P]_{(2)} \hx^2 =
  {\rm i} \frac{\beta}{f'(\langle X\rangle)} P^2,
\end{gather*}
from which we get the approximate commutation relation
\begin{gather*}
  [X, P] \simeq {\rm i} \biggl(1 + \frac{\beta}{f'(\langle X\rangle)} P^2\biggr).
\end{gather*}

We conclude that in those states for which $(\Delta X)^2 \sim \beta$, there is a nonzero
minimal uncertainty in $X$, given by $\Delta X_0 \simeq \sqrt{\beta/f'(\langle X
\rangle)}$ and therefore dependent on $\langle X\rangle$. For $f(X) = X$, such
 a~dependence disappears so that one retrieves Kempf's result, as it should be.

\subsection[Representations of $X$ and $P$ in terms of conventional position
and momentum operators]{Representations of $\boldsymbol{X}$ and $\boldsymbol{P}$ in terms of conventional position\\
and momentum operators}

In the case of Kempf's quadratic commutation relation corresponding to the choice $f(X)
= X$ in (\ref{eq:f-com}), several dif\/ferent kinds of representations of $X$ and $P$ in terms
of conventional opera\-tors~$x$,~$p$ have been used. Among them, one may quote the
momentum representation $P = p$, $X = (1 + \beta p^2) x$ (with $x = {\rm i}
d/dp$)~\cite{kempf95, kempf97, chang02b} and the (quasi)position representation $X
\simeq x$, $P \simeq p(1 + ({1}/{3}) \beta p^2)$ (with $p = - {\rm i}
d/dx$)~\cite{brau99}. Whereas the former is exact, the latter is only valid to f\/irst order in
$\beta$. As we now plan to show, the last one can be extended to the generalized
commutation relation (\ref{eq:f-com}).

Let us indeed look for a representation of (\ref{eq:f-com}) of the type
\begin{gather}
  X \simeq x, \qquad P \simeq p + \beta A(x,p), \label{eq:rep1}
\end{gather}
where $A(x,p)$ is a so far undetermined function of $x$ and $p$. Inserting this ansatz in
(\ref{eq:f-com}), we f\/ind that to f\/irst order in $\beta$, the latter is equivalent to $[f(x),
A(x,p)] = {\rm i} p^2$. Such a condition can be easily fulf\/illed by an operator $A$ such as
\begin{gather}
  A(x,p) = \frac{1}{2} \{\lambda(x), p\} + \frac{1}{2} \{\mu(x), p^3\}, \label{eq:A}
\end{gather}
where
\begin{gather}
  \mu(x) = \frac{1}{3f'}, \label{eq:mu}
  \\
  \lambda(x) = \mu \biggl[\frac{3}{2} \biggl(\frac{\mu'}{\mu}\biggr)'
  + \mu\biggl(\frac{1}{\mu}\biggr)''\biggr]
  = \frac{1}{6f'}\biggl(- \frac{f'''}{f'} + 3 \frac{f^{\prime\prime2}}{f^{\prime2}}\biggr),\label{eq:lambda}
\end{gather}
and a prime denotes derivative with respect to $x$. A further simplif\/ication occurs for the
choice made in (\ref{eq:f'}) since equation (\ref{eq:lambda}) can then be transformed
into
\begin{gather*}
  \lambda(x) = a + (b^2 - 4ac) \mu(x)
\end{gather*}
with $\mu(x)$ given in (\ref{eq:mu}).

{\sloppy
The representation (\ref{eq:rep1}) also allows us to calculate $[X, P]$ to f\/irst order in
$\beta$. From $[x, A(x,p)] = ({\rm i}/{2}) [2\lambda(x) + 3 \{\mu(x), p^2\}]$, we
indeed obtain
\begin{gather*}
  [X, P] \simeq {\rm i}\biggl[1 + \beta \biggl(\lambda(X) + \frac{3}{2} \{\mu(X), P^2\}
  \biggr)\biggr]
\end{gather*}
after substituting $X$ and $P$ for $x$ and $p$ in the f\/irst-order term on the right-hand
side.

}

For the examples considered in  (\ref{eq:f-ex1})--(\ref{eq:f-ex3}), the representation of
$P$ and the commutation relation $[X, P]$ are given by
\begin{gather}
  P  \simeq p + \frac{1}{6} \beta \{e^x, p + p^3\}, \nonumber \\
  [X, P]  \simeq {\rm i} \biggl[1 + \beta \biggl(\frac{1}{3} e^X + \frac{1}{2}\{e^X,
       P^2\}\biggr)\biggr], \label{eq:rep1-Morse} \\
  P  \simeq p + \beta\biggl(- p + \frac{1}{6} \{\cosh^2 x, 4p + p^3\}\biggr), \nonumber \\
  [X, P]  \simeq {\rm i} \biggl[1 + \beta \biggl(- 1 + \frac{4}{3} \cosh^2 X +
       \frac{1}{2}\{\cosh^2 X, P^2\}\biggr)\biggr], \label{eq:rep1-PT} \\
  P  \simeq p + \beta\biggl(p + \frac{1}{6} \{\cos^2 x, - 4p + p^3\}\biggr), \nonumber \\
  [X, P]  \simeq {\rm i} \biggl[1 + \beta \biggl(1 - \frac{4}{3} \cos^2 X +
       \frac{1}{2}\{\cos^2 X, P^2\}\biggr)\biggr],  \label{eq:rep1-PTbis}
\end{gather}
respectively.

For the special case of (\ref{eq:f-ex1}), it is worth mentioning the existence of an
alternative exact representation of $f(X)$ and $P$,
\begin{gather}
  -e^{-X} = - e^{-x} + \beta p^2, \qquad P = p. \label{eq:rep2}
\end{gather}

In view of the applications to quantum mechanics to be carried out in the next
subsection, it is worth inquiring into the restrictions, if any, that the Hermiticity of the
deformed operators~$X$ and $P$ imposes on square-integrable functions. For
representation (\ref{eq:rep1}), let us consider two functions $\psi(x), \phi(x) \in
L^2(x_1, x_2)$ and expand them to f\/irst order in $\beta$ as $\psi(x) = \psi^0(x) +
\beta \Delta \psi(x)$, $\phi(x) = \phi^0(x) + \beta \Delta \phi(x)$. Then, in the same
approximation, the Hermiticity of $P$ in $L^2(x_1, x_2)$ is equivalent to the condition
\begin{gather}
  \int_{x_1}^{x_2} dx\, \psi^{0*}(x) A(x,p) \phi^0(x) = \Biggl[\int_{x_1}^{x_2} dx\,
  \phi^{0*}(x) A(x,p) \psi^0(x)\Biggr]^*. \label{eq:cond-A}
\end{gather}
On using (\ref{eq:A}) and integrating several times by parts, it is easy to show that
equation (\ref{eq:cond-A}) amounts to the two restrictions
\begin{gather}
  \lambda |\psi^0|^2 \to 0 \qquad \text{\rm for\ } x \to x_1 \text{\ \rm and\ }
        x \to x_2, \label{eq:C1}
\\
2 \mu'' |\psi^0|^2 + \mu' (\psi^{0*} \psi^{0\prime} + \psi^{0\prime*} \psi^{0})
+ 2 \mu (\psi^{0*} \psi^{0\prime\prime} - |\psi^{0\prime*}|^2 + \psi^{0\prime
\prime*} \psi^0) \to 0 \nonumber
\\ \qquad
\text{\rm for\ } x \to x_1 \text{\ \rm and\ } x \to x_2,
\label{eq:C2}
\end{gather}
in the special case where $\phi^0(x) = \psi^0(x)$. In contrast, for representation
(\ref{eq:rep2}), no further condition is found to ensure the Hermiticity of $e^{-X}$.

\subsection{Applications to quantum mechanics}

Let us a consider a Hamiltonian
\begin{gather}
  H = \frac{1}{2} P^2 + V(X), \label{eq:H-f}
\end{gather}
where $X$ and $P$ satisfy the function-dependent commutation relation
(\ref{eq:f-com}) for some choice of~$f(X)$. To be able to solve the eigenvalue problem
for such a Hamiltonian, we shall impose that (i)~it is factorizable as shown in
(\ref{eq:fact-H}), where the operators $B^{\pm}$, now depending only on three real
parameters $g$, $s$, $r$, are chosen in the form
\begin{gather}
  B^{\pm}(g, s, r) = \frac{1}{\sqrt{2}}[\mp \,{\rm i}gP + s f(X) + r], \label{eq:B-f}
\end{gather}
and (ii) $H$ is the f\/irst member $H_0$ of a hierarchy of Hamiltonians $H_i$, $i=0, 1,
2,\ldots$, similar to~(\ref{eq:H-f}) and satisfying a SI condition of type (\ref{eq:SI}).

The factorization of $H$ implies that there exists a close connection between $f(X)$ and
$V(X)$, while the SI condition leads to equation (\ref{eq:f'}), expressing $f'(X)$ as a
second-degree polynomial in $f(X)$. The proof of these assertions relies on the relation
\begin{gather}
  B^{\pm}(g, s, r) B^{\mp}(g, s, r) = \frac{1}{2}\{g (g \mp \beta s) P^2 + [s f(X) +
  r]^2 \mp gs f'(X)\}  \label{eq:BB}
\end{gather}
resulting from (\ref{eq:f-com}) and (\ref{eq:B-f}). On comparing (\ref{eq:H-f}) with
(\ref{eq:BB}), it is indeed obvious that $V(X)$ can be written in terms of $f(X)$ and
$f'(X)$. On the other hand, since we need four equations (but not more) to determine
the four types of parameters $g_i$, $s_i$, $r_i$, and $\epsilon_i$ from the SI condition,
it is evident that $f'(X)$ should contain the same kind of terms as $[s f(X) + r]^2$, which
is achieved by assuming condition (\ref{eq:f'}) (see \cite{bagchi05} for a systematic use of
such a type of reasoning). As a~result, the precise relationship between $V(X)$ and
$f(X)$ reads
\begin{gather}
  V(X) = \frac{1}{2}[s(s - ag) f^2(X) + s(2r - bg) f(X) + r^2 - cgs] + \epsilon_0.
  \label{eq:V-f}
\end{gather}
Observe that in Kempf's case, $f(X) = X$ leads to a harmonic oscillator potential
\cite{cq03}.

When using the representation (\ref{eq:rep1}) of $X$ and $P$ valid to f\/irst order in
$\beta$, the Hamilto\-nian~(\ref{eq:H-f}) can be written in the same approximation as
\begin{gather*}
  H \simeq \frac{1}{2} p \biggl[1 + \frac{1}{6} \beta \biggl(2a - \frac{b^2-4ac}{f'(x)}
  \biggr)\biggr] p + \beta p^2 \frac{1}{3f'(x)} p^2 + V(x)
\end{gather*}
in terms of the operators $x$, $p$. As compared with conventional
Hamiltonians, there is a small additional term proportional to $p^4$ as in Kempf's
case~\cite{brau99}, but for $f(X) \ne X$ we also observe a position dependence of the
terms in $p^2$ and $p^4$. Furthermore, bound-state wavefunctions of~$H$ have not only
to be square integrable on the (f\/inite or inf\/inite) interval of def\/inition $(x_1, x_2)$
of~$V(x)$, but also to satisfy some additional restrictions ensuring the Hermiticity of $H$.
For such a~purpose, it is enough to impose that $P$ be Hermitian, hence that conditions
(\ref{eq:C1}) and (\ref{eq:C2}) be fulf\/illed.

We shall now proceed to consider several examples associated with the functions
(\ref{eq:f-ex1})--(\ref{eq:f-ex3}).

\subsubsection[P\"oschl-Teller potentials]{P\"oschl--Teller potentials}

For $f(X) = \tanh X$ corresponding to (\ref{eq:f-ex2}), we get the one-parameter
hyperbolic P\"oschl--Teller potential
\begin{gather*}
  V(X) = - \frac{1}{2} A(A+1) \sech^2 X, \qquad A \ge 1,
\end{gather*}
by choosing
\begin{gather}
  s = \sqrt{\frac{A(A+1)}{1+k}}, \qquad g = ks, \qquad r = 0, \qquad \epsilon_0 = -
  \frac{1}{2} s^2 \label{eq:s-PT}
\end{gather}
in (\ref{eq:V-f}). Here $k$ is def\/ined by
\begin{gather}
  k = \frac{1 + \beta A(A+1) + \Delta}{2A(A+1)}, \qquad \Delta \equiv \sqrt{[1 + \beta
  A(A+1)]^2 + 4A(A+1)}. \label{eq:k-PT}
\end{gather}
In the $\beta \to 0$ limit, equations (\ref{eq:s-PT}) and (\ref{eq:k-PT}) lead to the
conventional result $s \to A$, $g \to 1$, so that $B^{\pm}(g,s) \to B^{\pm}(A) =
(\mp {\rm i} p + A \tanh x)/\sqrt{2}$~\cite{cooper}.

The SI condition is then solved by considering the combinations of parameters
\begin{gather}
  u_i = g_i + {\rm i} \sqrt{\beta}\, s_i = |u_i| e^{{\rm i}\varphi_i}. \label{eq:u-PT}
\end{gather}
The results read
\begin{gather*}
  |u_i| = |u|, \qquad \varphi_i = \varphi + \frac{\rm i}{2} \phi, \qquad r_i = 0, \qquad
  \epsilon_{i+1} = \frac{1}{2}(s_i^2 - s_{i+1}^2),
\end{gather*}
where it follows from (\ref{eq:s-PT}) that
\begin{gather*}
  |u| = \sqrt{1 + \beta A(A+1)}, \qquad \cos\varphi = \frac{g}{\sqrt{1 + \beta A(A+1)}},
  \qquad \sin\varphi = \frac{\sqrt{\beta}s}{\sqrt{1 + \beta A(A+1)}},
\end{gather*}
and $\phi$ is def\/ined by
\begin{gather}
  e^{{\rm i}\phi} = \frac{1 - {\rm i}\sqrt{\beta}}{1 + {\rm i}\sqrt{\beta}}, \qquad
  \cos\phi = \frac{1 - \beta}{1 + \beta}, \qquad \sin\phi = - \frac{2\sqrt{\beta}}{1 +
  \beta}. \label{eq:phi-PT}
\end{gather}

The energy eigenvalues of the (generalized) hyperbolic P\"oschl--Teller Hamiltonian can be
expressed as
\begin{gather*}
  E_n = \sum_{i=0}^n \epsilon_i = - \frac{1}{2} s_n^2 = - \frac{|u|^2}{2\beta} \sin^2
  \biggl(\varphi + \frac{n}{2} \phi\biggr).
\end{gather*}
On expanding them into powers of $\beta$ and keeping only the f\/irst two terms, we
obtain
\begin{gather}
  E_n  \simeq - \frac{1}{2} (A-n)^2 (1 - \beta \delta_n + \cdots), \nonumber \\
  \delta_n  = \frac{1}{(2A+1)(A-n)} \biggl\{A^2 + n\biggl[(n+1)A - \frac{n^2+2}{3}
        \biggr](2A+1)\biggr\}. \label{eq:E-PT}
\end{gather}
In the $\beta \to 0$ limit, we get back the bound-state spectrum of the conventional
P\"oschl--Teller Hamiltonian, $E_n^0 = - (1/2) (A-n)^2$, $n=0, 1, \ldots, n_{\max}$,
$A-1 \le n_{\max} < A$~\cite{cooper}, as it should be. It can be easily shown
that for $\beta \ne 0$, the f\/irst-order relative correction in (\ref{eq:E-PT}) has the
opposite sign and increases with $n$: $0 < \delta_0 < \delta_1 < \cdots <
\delta_{n_{\max}}$. For the approximation (\ref{eq:E-PT}) to be meaningful, we must
therefore restrict ourselves to $\beta$ values such that $\beta \ll (\delta_{n_{\max}})^{-1}$,
which for integer $A$, for instance, gives rise to the condition $\beta \ll 3
(2A+1) (4A^4 + 2A^3 - 7A^2 + A + 3)^{-1}$.

We can get the corresponding eigenfunctions
\begin{gather*}
  \psi_n(g,s;x) \simeq \psi_n^0(A;x) + \beta \Delta \psi_n(g,s;x)
\end{gather*}
in the representation (\ref{eq:rep1-PT}), wherein the (generalized) P\"oschl--Teller
Hamiltonian and the opera\-tors $B^{\pm}(g,s)$ take the form
\begin{gather}
  H \simeq \frac{1}{2} p \biggl[1 - \frac{1}{3} \beta (1 + 2 \cosh^2 x)\biggr] p +
      \frac{1}{3} \beta p^2 \cosh^2 x\, p^2 - \frac{1}{2} A(A+1) \sech^2 x,
      \label{eq:H-PT}
\\
      B^{\pm}(g,s)  \simeq B^{\pm}_0(A) + \beta \Delta B^{\pm}(g,s), 
      \nonumber
\\ \Delta B^{\pm}(g,s)  = \frac{1}{\sqrt{2}} \biggl[\mp \frac{\rm i}{2} (A-2) p  \mp
      \frac{\rm i}{6} \{\cosh^2 x, 4p + p^3\} - \frac{A^2}{2(2A+1)} \tanh x \biggr],
\end{gather}
with $p = - {\rm i} d/dx$.

Since the ground state wavefunction $\psi_0(g,s;x)$ must be annihilated by the operator
$B^-(g,s)$, the f\/irst-order correction term $\Delta \psi_0(g,s;x)$ with respect to
$\psi_0^0(A;x) = N_0(A) \sech^A x$ (where $N_0(A) =
\{\Gamma(A+{1}/{2})/[\Gamma({1}/{2}) \Gamma(A)]\}^{1/2}$) satisf\/ies the
f\/irst-order dif\/ferential equation
\begin{gather*}
  B^-_0(A) \Delta \psi_0(g,s;x) = - \Delta B^-(g,s) \psi_0^0(A;x).
\end{gather*}
A straightforward calculation leads to
\begin{gather*}
  \Delta \psi_0(g,s;x) = N_0(A) \biggl[- \frac{1}{6} A(A-1)(A-2) \sech^{A-2} x + C
  \sech^A x \ln\cosh x + D \sech^A x\biggr],
\end{gather*}
with $C \equiv A(A+1)(2A^2 + 2A - 1)/[3(2A+1)]$, while $D$ is some integration
constant. The latter can be determined from the normalization condition of
$\psi_0(g,s;x)$ to f\/irst order in $\beta$. In the special case where $A$ is integer, it is
given by
\begin{gather*}
  D = \frac{1}{12} A(A-1)(2A-1) + C \biggl(\ln2 + \sum_{j=1}^{2A-1} \frac{(-1)^j}{j}
  \biggr).
\end{gather*}
Finally, it can be easily shown that for $\psi_0(g,s;x)$, the conditions (\ref{eq:C1}) and
(\ref{eq:C2}) imposed by the Hermiticity of $P$ (hence of $H$) amount to the
restriction $\sech^{2A-2} x \to 0$ for $x \to \pm \infty$. The allowed $A$ values are
therefore $A >1$ (excluding $A=1$).

The excited-state wavefunctions can be calculated in a similar way. For instance, the
f\/irst-order correction $\Delta \psi_1(g,s;x)$ to the f\/irst-excited state wavefunction can
be obtained from
\begin{gather*}
  \Delta \psi_1(g,s;x) \propto B^+_0(A) \Delta \psi_0(g_1,s_1;x) + \Delta B^+(g,s)
  \psi_0^0(A-1;x),
\end{gather*}
where $g_1$ and $s_1$ are given by equations (\ref{eq:u-PT})--(\ref{eq:phi-PT}).

The treatment of the one-parameter trigonometric P\"oschl--Teller potential
\begin{gather*}
  V(X) = \frac{1}{2} A(A-1) \sec^2 X, \qquad - \frac{\pi}{2} \le X \le \frac{\pi}{2},
  \qquad A > 1,
\end{gather*}
is entirely analogous. With $f(X) = \tan X$ corresponding to (\ref{eq:f-ex3}) and
(\ref{eq:rep1-PTbis}), we are led to the energy spectrum
\begin{gather}
  E_n \simeq \frac{1}{2} (A+n)^2 (1 + \beta \delta_n + \cdots), \nonumber
  \\
  \delta_n = \frac{1}{(2A-1)(A+n)} \biggl\{A^2 + n\biggl[(n+1)A + \frac{n^2+2}{3}
        \biggr](2A-1)\biggr\},
\end{gather}
valid up to f\/irst order in $\beta$. As previously, the correction terms satisfy the property
$0 < \delta_0 < \delta_1 < \cdots < \delta_n < \cdots$.

\subsubsection{Morse potential}

The Morse potential
\begin{gather*}
  V(X) = \frac{1}{2} B^2 e^{-2X} - \frac{1}{2} B (2A+1) e^{-X}, \qquad A, B > 0,
\end{gather*}
is associated with the choice $f(X) = - e^{-X}$ in (\ref{eq:f-ex1}). In such a case, it is
straightforward to show that
\begin{gather}
  s  = B, \qquad g = \frac{1}{2} \beta B + \Delta, \qquad r = A + \frac{1}{2} -
      \frac{g}{2}, \qquad \Delta \equiv \sqrt{1 + \frac{1}{4} \beta^2 B^2}, \nonumber \\
  s_i  = s, \qquad g_i = g + i \beta B, \qquad r_i = r - ig - \frac{1}{2} i^2 \beta B,
      \label{eq:s-Morse}
\end{gather}
and
\begin{gather}
  E_n = - \frac{1}{2} r_n^2 = - \frac{1}{2} \biggl[A + \frac{1}{2} - \frac{1}{4} (2n^2 +
  2n + 1) \beta B - \biggl(n + \frac{1}{2}\biggr) \Delta\biggr]^2, \label{eq:E-Morse}
\end{gather}
where in the $\beta \to 0$ limit, we obtain the conventional energy spectrum $E_n^0 =
- ({1}/{2})(A-n)^2$, $n=0, 1, \ldots,n_{\max}$, $A-1 \le n_{\max} < A$
\cite{cooper}.

First-order corrections to the corresponding wavefunctions can be obtained by using the
representation (\ref{eq:rep1-Morse}) as for P\"oschl--Teller potentials. We shall instead
use here the alternative representation (\ref{eq:rep2}) because it leads to exact results
for the Hamiltonian
\begin{gather}
  H = \frac{1}{2} p\, [1 \!+\! \beta B(2A \!+\! 1\! - \!2B e^{-x})] p +\! \frac{1}{2} \beta^2 B^2 p^4
  +\! \frac{1}{2} B^2 e^{-2x} -\! \frac{1}{2} B (2A + 1 - \beta B) e^{-x}. \label{eq:H-Morse}
\end{gather}
As observed at the end of Section 3.2, bound-state wavefunctions of $H$ do not have to
satisfy any extra condition apart from square integrability on the real line.

As shown in the appendix, by employing the properties of the operators
\begin{gather}
  B^{\pm}(g,B,r) = \frac{1}{\sqrt{2}} \biggl(- \beta B \frac{d^2}{dx^2} \mp g
  \frac{d}{dx} - B e^{-x} + r\biggr), \label{eq:B-Morse}
\end{gather}
the ground- and excited-state wavefunctions of such a Hamiltonian, corresponding to the
energy eigenvalues (\ref{eq:E-Morse}), can be expressed in terms of Bessel functions as
\begin{gather}
  \psi_0(g,B,r;x) \propto \exp\biggl(\frac{gx}{2\beta B}\biggr) J_{\nu}\biggl(
  \frac{2}{\sqrt{\beta}} e^{-x/2}\biggr) \label{eq:gs-Morse}
\end{gather}
and
\begin{gather}
  \psi_n(g,B,r;x) \propto \sum_{j=0}^n c_j \exp\biggl(\frac{g_{n-j}x}{2\beta B}\biggr)
  J_{\nu_n+j}\biggl( \frac{2}{\sqrt{\beta}} e^{-x/2}\biggr), \label{eq:es-Morse}
\end{gather}
respectively. Here $\nu$ is def\/ined by
\begin{gather}
  \nu = \frac{1}{\beta B} \sqrt{g^2 + 4\beta Br} \label{eq:nu}
\end{gather}
and a similar expression applies to $\nu_n$ in terms of $g_n$ and $r_n$, while $c_j$
denote some constants.

From the properties of Bessel functions \cite{abramowitz}, it follows that for $x \to -
\infty$, $\psi_0(g,B,r;x)$ decays exponentially as $\exp[-({g}/({2\beta B})+{1}/{4}) |x|]$,
while for $x \to \infty$, it behaves as $\exp[({g}/(2\beta B) - {\nu}/{2}) x]$.
It is therefore normalizable on the
real line provided $\nu > g/(\beta B)$ or, in other words, $r > 0$. From the expression of
$r$ given in (\ref{eq:s-Morse}), we conclude that the Hamiltonian (\ref{eq:H-Morse}) has
at least one bound state, given by (\ref{eq:gs-Morse}), if $\beta$ takes a value satisfying
the inequality $\beta < 4A(A+1) [(2A+1)B]^{-1}$. Whenever such a condition is
satisf\/ied, there may actually also exist excited states. Since the normalizability condition
of $\psi_n(g,B,r;x)$, given in (\ref{eq:es-Morse}), is $r_n > 0$, the allowed values of $n$
are $n=0, 1, \ldots, n_{\max}$, where $n_{\max}$ is the largest integer for
which $r_n > 0$, i.e., the integer fulf\/illing the condition
\begin{gather}
\frac{1}{\beta B} \biggl(\!\!- \frac{3}{2} \beta B  \!-\! \Delta \!+\! \sqrt{1 \!+\! \beta B(2A\!+\!1)}
  \biggr)\! \le\! n_{\max}< \! \frac{1}{\beta B} \biggl(\!\!- \frac{1}{2} \beta B \! -\! \Delta \!+\!
  \sqrt{1 + \beta B(2A\!+\!1)} \biggr).
\label{eq:number -Morse}
\end{gather}
Observe that for $\beta \to 0$, equation (\ref{eq:number -Morse}) leads to the
conventional result $A-1 \le n_{\max} < A$, as it should be.

\section{Final remarks}

In the present paper, we have considered two extensions of Kempf's quadratic canonical
commutation relation in one dimension. For both of them, we have studied how they alter
the minimal uncertainties in position and/or momentum and we have proposed some
applications to quantum mechanics.

The f\/irst one is the most general quadratic commutation relation, which turns out to also
include a $q$-deformed Heisenberg algebra as a special case. Although it can be reduced
to Kempf's commutation relation by applying some transformations to the position and
momentum operators, we have shown that it gives rise to more general expressions for
the nonzero minimal uncertainties in position and momentum and that it has a def\/inite
inf\/luence on the energy spectrum and the corresponding eigenfunctions of the harmonic
oscillator in an electric f\/ield, f\/irst considered in \cite{cq04a}.

In the second generalization, the position operator $X$ has been replaced in the simplest
quadratic commutation relation (with only a nonzero minimal uncertainty in position) by
some function $f(X)$ in such a way that the choice $f(X) = X$ gives back the original
commutation relation. From the uncertainty point of view, there appears a new and
interesting ef\/fect in the sense that the minimal uncertainty in $X$ now becomes
dependent on the average position $\langle X\rangle$ through the function $f'(\langle
X\rangle)$.

Next, to any function-dependent commutation relation of this type, such that $f'(X)$
satisf\/ies equation (\ref{eq:f'}), we have associated a def\/inite family of potentials
$V(X)$, whose spectrum can be exactly determined through SUSYQM and SI techniques
in the same way as we had done for the harmonic oscillator in the case of the original
choice $f(X) = X$ \cite{cq03}. Along these lines, we have treated in detail the cases
of the one-parameter hyperbolic and trigonometric P\"oschl--Teller potentials, as well as
that of the Morse potential.

Furthermore, for any $f(X)$ we have obtained a representation of our deformed algebra
in terms of conventional position and momentum operators $x$, $p$, up to f\/irst order in
the deformation parameter $\beta$. This has allowed us to write the corresponding
Hamiltonian in terms of $x$, $p$, and to determine its bound-state wavefunctions in the
same approximation. Finally, for the Morse potential, the use of an alternative exact
representation of the associated algebra has led us to some exact results for the
Hamiltonian and its bound-state wavefunctions.

In both types of representations, the resulting Hamiltonians are of fourth order in $p$
and, in addition, the $p$-dependent terms may also depend on $x$. The former property
had already been observed elsewhere \cite{brau99} in the context of Kempf's quadratic
commutation relations, while the latter reminds one of the well-known equivalence
between some deformed commutation relations and position-dependent masses (PDM)
\cite{cq04b}.

Both characteristics may be very interesting in the area of condensed-matter problems.
A~dependence of the kinetic energy on the position is indeed often used there due to its
relevance in describing the dynamics of electrons in compositionally-graded crystals
\cite{geller}, quantum dots \cite{serra} and liquid crystals \cite{barranco}, for instance.
Furthermore, it has been shown that fourth-order terms in the momentum may depict
nonparabolicity ef\/fects in quantum wells \cite{ekenberg}.

Some applications may arise in other f\/ields too. Let us mention the occurrence of PDM in
the energy-density functional approach to the quantum many-body problem in the
context of nonlocal terms of the accompanying potential with applications to nuclei
\cite{ring}, quantum li\-quids~\cite{arias} and metal clusters \cite{puente}. It is also worth
noting that higher-order terms in $p^2$ appear in semiclassical approaches to the
Klein-Gordon equation, as well as in Hermitian Hamiltonians equivalent to $\cal
PT$-symmetric or, more generally, pseudo-Hermitian ones \cite{jones, mosta, bagchi06}.

In conclusion, we would like to stress the relevance of the innovative deformed-algebra
approach proposed here to derive exact or approximate solutions to Schr\"odinger
equations containing generalized Hamiltonians such as those given in (\ref{eq:H-PT}) and
(\ref{eq:H-Morse}). We do think that it will open a new inspiration for future study.

\section*{Appendix. Bound-state wavefunctions of the generalized\\ Morse Hamiltonian
(\ref{eq:H-Morse})}
\renewcommand{\theequation}{A.\arabic{gather}}

To calculate the bound-state wavefunctions $\psi_n(g,B,r;x)$ of the Hamiltonian
(\ref{eq:H-Morse}), it proves convenient to introduce two auxiliary variables
\begin{gather}
  y = B e^{-x}, \qquad z = 2 \sqrt{\frac{y}{\beta B}}, \tag{A.1}\label{eq:variables}
\end{gather}
both varying on the half line $(0, \infty)$, and to def\/ine
\begin{gather}
  \psi_n(g,B,r;x) = \phi_n(g,B,r;y) = \chi_n(g,B,r;z), \tag{A.2}\label{eq:functions}
\end{gather}
satisfying the normalization condition
\begin{gather}
  \int_{-\infty}^{+\infty}\! dx\, |\psi_n(g,B,r;x)|^2\! =\! \int_0^{+\infty} \frac{dy}{y}\,
  |\phi_n(g,B,r;y)|^2 \!=\! 2 \int_0^{+\infty} \frac{dz}{z}\, |\chi_n(g,B,r;z)|^2\! =\! 1.\tag{A.3}
  \label{eq:norm}
\end{gather}
The operators $B^{\pm}(g,B,r)$ in equation (\ref{eq:B-Morse}) can then be written as
\begin{gather*}
  B^{\pm}(g,B,r) = \frac{1}{\sqrt{2}} \biggl[- \beta B \biggl(y \frac{d}{dy}\biggr)^2 \pm
  g y \frac{d}{dy} - y + r\biggr].
\end{gather*}

{\samepage
The ground-state wavefunction being a zero mode of $B^-(g,B,r)$ is the solution of the
second-order dif\/ferential equation
\begin{gather}
  \biggl[\beta B y^2 \frac{d^2}{dy^2} + (g + \beta B) y \frac{d}{dy} + y - r\biggr]
  \phi_0(g,B,r;y) = 0, \tag{A.4}\label{eq:diff-eqn}
\end{gather}
satisfying condition (\ref{eq:norm}). Since $y=0$ is a regular singular point of this
equation, we may look for a solution of the type
\begin{gather}
  \phi_0(g,B,r;y) = y^{\rho} \sum_{j=0}^{\infty} a_j y^j, \tag{A.5}\label{eq:ansatz}
\end{gather}
where $a_j$ are some constants, $a_0 \ne 0$ and $\rho > 0$ in order to fulf\/il
(\ref{eq:norm}). Inserting (\ref{eq:ansatz}) in (\ref{eq:diff-eqn}), we easily f\/ind that
$\rho = ({1}/{2})(- {g}/(\beta B) + \nu)$ (with $\nu$ def\/ined in
(\ref{eq:nu})), provided the condition $r > 0$ is satisf\/ied. Furthermore
\begin{gather*}
  \sum_{j=0}^{\infty} a_j y^j \propto {}_0F_1\biggl(\nu + 1; - \frac{y}{\beta B}\biggr)
  \propto y^{-\nu/2} J_{\nu}(z),
\end{gather*}
where in the last step use is made of equation (9.1.69) of \cite{abramowitz}. This proves
equation (\ref{eq:gs-Morse}).

}

{\sloppy
The f\/irst-excited state wavefunction can be obtained by acting with $B^+(g,B,r)$ on
$\psi_0(g_1,B,r_1;x)$. On using the analogue of equation (\ref{eq:diff-eqn}) for
$\phi_0(g_1,B,r_1;y)$ to eliminate the second-order derivative with respect to $y$, it
can be shown that
\begin{gather*}
  \phi_1(g,B,r;y) \propto \biggl[(g + g_1) y \frac{d}{dy} + r - r_1\biggr]
  \phi_0(g_1,B,r_1;y).
\end{gather*}
Since from (\ref{eq:s-Morse}) it results that $r - r_1 = ({1}/{2})(g + g_1)$, this
equation reduces to
\begin{gather*}
  \chi_1(g,B,r;z) \propto \biggl(z \frac{d}{dz} + 1\biggr) \chi_0(g_1,B,r_1;z).
\end{gather*}
Equation (9.1.29) of \cite{abramowitz} then leads to
\begin{gather*}
  \chi_1(g,B,r;z) \propto - z^{-g/(\beta B)} J_{\nu_1+1}(z) + (2\rho_1 + 1)
  z^{-g_1/(\beta B)} J_{\nu_1}(z)
\end{gather*}
with $\rho_1 = ({1}/{2})(- {g_1}/({\beta B}) + \nu_1)$. Combining this
result with (\ref{eq:variables}) and (\ref{eq:functions}) f\/inally yields equation
(\ref{eq:es-Morse}) for $n=1$.

}

The proof of such an equation for higher $n$ values is based upon observing that
\begin{gather}
  \phi_n(g,B,r;y) \propto \Biggl\{\prod_{j=0}^{n-1} \biggl[(g_j + g_n)\biggl(y \frac{d}{dy}
  - j\biggr) + r_j - r_n\biggr]\Biggr\} \phi_0(g_n,B,r_n;y) \tag{A.6}\label{eq:phi_n}
\end{gather}
and applying $n$ times equation (9.1.29) of \cite{abramowitz}. Equation
(\ref{eq:phi_n}) itself can be demonstrated by induction over $n$ by starting from the
relation $\phi_n(g,B,r;y) \propto B^+(g,B,r) \phi_{n-1}(g_1,B,r_1;y)$, commuting
$B^+(g,B,r)$ with the factors on its right until it acts on $\phi_0(g_n,B,r_n;y)$ and
eliminating the second-order derivative as done above.

\subsection*{Acknowledgements}

V.M.T.~thanks the National Fund for Scientif\/ic Research (FNRS), Belgium, for f\/inancial
support.

\pdfbookmark[1]{References}{ref}
\LastPageEnding


\begin{thebibliography}{99}

\footnotesize\itemsep=0pt

\bibitem{gross} Gross D.J., Mende P.F., String theory beyond the Planck scale,
{\it Nuclear Phys. B} \textbf{303} (1988), 407--454.

\bibitem{amati} Amati D., Ciafaloni M., Veneziano G., Can spacetime be probed below the
string size?, {\it Phys.\ Lett.\ B} \textbf{216} (1989), 41--47.

\bibitem{maggiore} Maggiore M., The algebraic structure of the generalized uncertainty
principle, {\it Phys.\ Lett.\ B} \textbf{319} (1993), 83--86,
\href{http://arxiv.org/abs/hep-th/9309034}{hep-th/9309034}.

\bibitem{connes} Connes A., Gravity coupled with matter and the foundation of
non-commutative geometry, {\it Comm. Math. Phys.} \textbf{182} (1996), 155--176,
\href{http://arxiv.org/abs/hep-th/9603053}{hep-th/9603053}.

\bibitem{amelino} Amelino-Camelia G., Mavromatos N.E., Ellis J., Nanopoulos D.V., On the
space-time uncertainty relations of Liouville strings and D-branes,
{\it Modern Phys. Lett. A} \textbf{12} (1997), 2029--2036,
\href{http://arxiv.org/abs/hep-th/9701144}{hep-th/9701144}.

\bibitem{seiberg} Seiberg N., Witten E., String theory and noncommutative geometry,
{\it JHEP} {\bf 9909} (1999), 032, 93 pages, \href{http://arxiv.org/abs/hep-th/9908142}{hep-th/9908142}.

\bibitem{kempf94a} Kempf A., Uncertainty relation in quantum mechanics with quantum group symmetry,
{\it J.\ Math.\ Phys.} \textbf{35} (1994), 4483--4496, \href{http://arxiv.org/abs/hep-th/9311147}{hep-th/9311147}.

\bibitem{hinrichsen} Hinrichsen H., Kempf A., Maximal localization in the presence of minimal
uncertainties in positions and in momenta, {\it J.\ Math.\ Phys.} \textbf{37} (1996), 2121--2137,
\href{http://arxiv.org/abs/hep-th/9510144}{hep-th/9510144}.

\bibitem{kempf95} Kempf A., Mangano G., Mann R.B., Hilbert space representation of the
minimal length uncertainty relation, {\it Phys.\ Rev.\ D} \textbf{52} (1995), 1108--1118,
\href{http://arxiv.org/abs/hep-th/9412167}{hep-th/9412167}.

\bibitem{kempf97} Kempf A., Non-pointlike particles in harmonic oscillators,
{\it J. Phys. A: Math. Gen.} \textbf{30} (1997), 2093--2102, \href{http://arxiv.org/abs/hep-th/9604045}{hep-th/9604045}.

\bibitem{kempf94b} Kempf A., Quantum f\/ield theory with nonzero minimal
uncertainties in position and momentum, \mbox{\href{http://arxiv.org/abs/hep-th/9405067}{hep-th/9405067}.}

\bibitem{brau99} Brau F., Minimal length uncertainty relation and the hydrogen atom,
{\it J. Phys. A: Math. Gen.} \textbf{32} (1999), 7691--7696, \href{http://arxiv.org/abs/quant-ph/9905033}{quant-ph/9905033}.

\bibitem{benczik05} Benczik S., Chang L.N., Minic D., Takeuchi T., Hydrogen-atom spectrum
under a minimal-length hypothesis, {\it Phys.\ Rev.\ A} \textbf{72} (2005), 012104, 4 pages,
\href{http://arxiv.org/abs/hep-th/0502222}{hep-th/0502222}.

\bibitem{stetsko} Stetsko M.M., Tkachuk V.M., Perturbation hydrogen-atom spectrum
in deformed space with minimal length, {\it Phys.\ Rev.\ A} \textbf{74} (2006), 012101, 5 pages.

\bibitem{brau06} Brau F., Buisseret F., Minimal length uncertainty relation and
gravitational quantum well, {\it Phys.\ Rev.\ D} \textbf{74} (2006), 036002, 5 pages,
\href{http://arxiv.org/abs/hep-th/0605183}{hep-th/0605183}.

\bibitem{nouicer} Nouicer K., An exact solution of the one-dimensional Dirac oscillator in the
presence of minimal lengths, {\it J.~Phys.~A: Math. Gen.} \textbf{39} (2006), 5125--5134.

\bibitem{chang02a} Chang L.N., Minic D., Okamura N., Takeuchi T., Ef\/fect of the minimal
length uncertainty relation on the density of states and the cosmological constant problem,
{\it Phys.\ Rev.\ D} \textbf{65} (2002), 125028, 7 pages, \href{http://arxiv.org/abs/hep-th/0201017}{hep-th/0201017}.

\bibitem{benczik02} Benczik S., Chang L.N., Minic D., Okamura N., Rayyan S., Takeuchi T.,
Short distance versus long distance physics: The classical limit of the minimal length
uncertainty relation, {\it Phys.\ Rev.\ D} \textbf{66} (2002), 026003, 11~pages,
\href{http://arxiv.org/abs/hep-th/0204049}{hep-th/0204049}.

\bibitem{chang02b} Chang L.N., Minic D., Okamura N., Takeuchi T., Exact solution of the
harmonic oscillator in arbit\-rary dimensions with minimal length uncertainty relations, {\it
Phys.\ Rev.\ D} \textbf{65} (2002), 125027, 8 pages, \href{http://arxiv.org/abs/hep-th/0111181}{\mbox{hep-th/0111181}}.

\bibitem{fityo06a} Fityo T.V., Vakarchuk I.O., Tkachuk V.M., One-dimensional Coulomb-like
problem in deformed space with minimal length, {\it J. Phys. A: Math. Gen.} \textbf{39} (2006),
2143--2150, \href{http://arxiv.org/abs/quant-ph/0507117}{quant-ph/0507117}.

\bibitem{fityo06b} Fityo T.V., Vakarchuk I.O., Tkachuk V.M., WKB approximation in deformed
space with minimal length, {\it J. Phys. A: Math. Gen.} \textbf{39} (2006), 379--388,
\href{http://arxiv.org/abs/quant-ph/0510018}{quant-ph/0510018}.

\bibitem{kempf93} Kempf A., Quantum group symmetric Bargmann--Fock space: integral
kernels, Green functions, driving forces, {\it J.\ Math.\ Phys.} \textbf{34} (1993), 969--987.

\bibitem{cq03} Quesne C., Tkachuk V.M., Harmonic oscillator with nonzero minimal
uncertainties in both position and momentum in a SUSYQM framework, {\it J. Phys. A: Math. Gen.} \textbf{36} (2003),
10373--10390, \href{http://arxiv.org/abs/math-ph/0306047}{math-ph/0306047}.

\bibitem{cooper} Cooper F., Khare A., Sukhatme U., Supersymmetry and quantum mechanics,
{\it Phys.\ Rep.} \textbf{251} (1995), 267--385, \href{http://arxiv.org/abs/hep-th/9405029}{hep-th/9405029}.

\bibitem{gendenshtein} Gendenshtein L.E., Derivation of exact spectra of the Schr\"odinger
equation by means of supersymmetry, {\it JETP Lett.} \textbf{38} (1983), 356--359.

\bibitem{dabrowska} Dabrowska J., Khare A., Sukhatme U., Explicit wavefunctions for
shape-invariant potentials by operator techniques, {\it J. Phys. A: Math. Gen.} \textbf{21} (1988), L195--L200.

\bibitem{spiridonov92a} Spiridonov V., Exactly solvable potentials and quantum algebras,
{\it Phys.\ Rev.\ Lett.} \textbf{69} (1992), 398--401, \href{http://arxiv.org/abs/hep-th/9112075}{hep-th/9112075}.

\bibitem{spiridonov92b} Spiridonov V., Deformed conformal and supersymmetric quantum
mechanics, {\it Modern Phys. Lett. A} \textbf{7} (1992), 1241--1252,
\href{http://arxiv.org/abs/hep-th/9202013}{hep-th/9202013}.

\bibitem{khare} Khare A., Sukhatme U.P., New shape-invariant potentials in supersymmetric
quantum mechanics, {\it J.~Phys.~A: Math. Gen.} \textbf{26} (1993), L901--L904,
\href{http://arxiv.org/abs/hep-th/9212147}{hep-th/9212147}.

\bibitem{barclay} Barclay D.T., Dutt R., Gangopadhyaya A., Khare A., Pagnamenta A.,
Sukhatme U., New exactly solvable Hamiltonians: shape invariance and self-similarity, {\it Phys.\
Rev.\ A} \textbf{48} (1993), 2786--2797, \href{http://arxiv.org/abs/hep-ph/9304313}{hep-ph/9304313}.

\bibitem{lutzenko} Lutzenko I., Spiridonov V., Zhedanov A., On the spectrum of a
$q$-oscillator with a linear interaction, {\it Phys.\ Lett.\ A} \textbf{204} (1995), 236--242.

\bibitem{loutsenko} Loutsenko I., Spiridonov V., Vinet L., Zhedanov A., Spectral analysis of
$q$-oscillator with general bilinear interaction, {\it J. Phys. A: Math. Gen.} \textbf{31} (1998),
9081--9094.

\bibitem{cq04a} Quesne C., Tkachuk V.M., More on a SUSYQM approach to the harmonic
oscillator with nonzero minimal uncertainties in position and/or momentum,
{\it J. Phys. A: Math. Gen.} \textbf{37} (2004), 10095--10114,
\href{http://arxiv.org/abs/math-ph/0312029}{\mbox{math-ph/0312029}}.

\bibitem{cq05} Quesne C., Tkachuk V.M., Dirac oscillator with nonzero minimal uncertainty in
position, {\it J. Phys. A: Math. Gen.} \textbf{38} (2005), 1747--1766,
\href{http://arxiv.org/abs/math-ph/0412052}{math-ph/0412052}.

\bibitem{pillin} Pillin M., On the deformability of Heisenberg algebras,
{\it Comm. Math. Phys.} \textbf{180} (1996), 23--38, \href{http://arxiv.org/abs/q-alg/9508014}{\mbox{q-alg/9508014}}.

\bibitem{bagchi05} Bagchi B., Banerjee A., Quesne C., Tkachuk V.M., Deformed shape invariance
and exactly solvable Hamiltonians with position-dependent ef\/fective mass,
{\it J. Phys. A: Math. Gen.} \textbf{38} (2005), 2929--2946,
\href{http://arxiv.org/abs/quant-ph/0412016}{\mbox{quant-ph/0412016}}.

\bibitem{abramowitz} Abramowitz M., Stegun I.A., Handbook of mathematical
functions, Dover, New York, 1965.

\bibitem{cq04b} Quesne C., Tkachuk V.M., Deformed algebras, position-dependent ef\/fective
masses and curved spaces: an exactly solvable Coulomb problem, {\it J. Phys. A: Math. Gen.} \textbf{37} (2004),
4267--4282, \href{http://arxiv.org/abs/math-ph/0403047}{math-ph/0403047}.

\bibitem{geller} Geller M.R., Kohn W., Quantum mechanics of electrons in crystals with graded
composition, {\it Phys.\ Rev.\ Lett.} \textbf{70} (1993), 3103--3106.

\bibitem{serra} Serra Ll., Lipparini E., Spin response of unpolarized quantum dots, {\it
Europhys.\ Lett.} \textbf{40} (1997), 667--672.

\bibitem{barranco} Barranco M., Pi M., Gatica S.M., Hern\'andez E.S., Navarro J., Structure and
energetics of mixed ${}^4$He-$^3$He drops, {\it Phys.\ Rev.\ B} \textbf{56} (1997), 8997--9003.

\bibitem{ekenberg} Ekenberg U., Nonparabolicity ef\/fects in a quantum well: sublevel shift,
parallel mass, and Landau levels, {\it Phys.\ Rev.\ B} \textbf{40} (1989), 7714--7726.

\bibitem{ring} Ring P., Schuck P., The nuclear many body problem, Springer, New York,
1980.

\bibitem{arias} Arias de Saavedra F., Boronat J., Polls A., Fabrocini A., Ef\/fective mass of one
${}^4$He atom in liquid ${}^3$He, {\it Phys.\ Rev.\ B} \textbf{50} (1994), 4248--4251.

\bibitem{puente} Puente A., Serra Ll., Casas M., Dipole excitation of Na clusters with a
non-local energy density functional, {\it Z.\ Phys.\ D} \textbf{31} (1994), 283--286.

\bibitem{jones} Jones H.F., On pseudo-Hermitian Hamiltonians and their Hermitian counterparts,
{\it J. Phys. A: Math. Gen.} \textbf{38} (2005), 1741--1746.

\bibitem{mosta} Mostafazadeh A., $\cal PT$-symmetric cubic anharmonic oscillator as a
physical model, {\it J. Phys. A: Math. Gen.} \textbf{38} (2005), 6557--6570,
Corrigendum, {\it J. Phys. A: Math. Gen.} \textbf{38} (2005), 8185--8185,
\href{http://arxiv.org/abs/quant-ph/0411137}{quant-ph/0411137}.

\bibitem{bagchi06} Bagchi B., Quesne C., Roychoudhury R., Pseudo-Hermitian versus Hermitian
position-dependent-mass Hamiltonians in a perturbative framework, {\it J. Phys. A: Math. Gen.} \textbf{39}
(2006), L127--L134, \href{http://arxiv.org/abs/quant-ph/0511182}{\mbox{quant-ph/0511182}}.

\end{thebibliography}
\end{document}